\pdfoutput=1

\documentclass[10pt]{article}

\usepackage{ACL2023}

\usepackage{times}
\usepackage{latexsym}

\usepackage{listings}

\usepackage{adjustbox}

\usepackage{hyperref}
\usepackage{multirow}
\usepackage[normalem]{ulem}
\useunder{\uline}{\ul}{}
\usepackage{algpseudocode}
\usepackage{algorithm}

\usepackage[T1]{fontenc}
\usepackage{longtable}

\usepackage{graphicx}
\usepackage{booktabs}

\usepackage[utf8]{inputenc}

\usepackage{amsmath}
\usepackage{listings}
\usepackage{xcolor}

\definecolor{codegreen}{rgb}{0,0.6,0}
\definecolor{codegray}{rgb}{0.5,0.5,0.5}
\definecolor{codepurple}{rgb}{0.58,0,0.82}
\definecolor{backcolour}{rgb}{0.95,0.95,0.92}

\lstdefinestyle{mystyle}{
    backgroundcolor=\color{backcolour},   
    commentstyle=\color{codegreen},
    keywordstyle=\color{magenta},
    numberstyle=\tiny\color{codegray},
    stringstyle=\color{codepurple},
    basicstyle=\ttfamily\footnotesize,
    breakatwhitespace=false,         
    breaklines=true,                 
    captionpos=b,                    
    keepspaces=true,                 
    numbers=left,                    
    numbersep=5pt,                  
    showspaces=false,                
    showstringspaces=false,
    showtabs=false,                  
    tabsize=2
}

\usepackage{microtype}

\usepackage{inconsolata}

\newcommand{\mono}[1]{{\footnotesize\texttt{#1}}}
\newcommand*\NOINDENT{\@@par   
      \@totalleftmargin\z@ \@listdepth\z@ \rightmargin\z@
}
\usepackage{float}

\title{Repository-level Code Search with Neural Retrieval Methods}

\author{Siddharth Gandhi, Luyu Gao, Jamie Callan \\
  School of Computer Science\\
  Carnegie Mellon University\\
  \texttt{\{ssg2,luyug,callan\}@cs.cmu.edu}}

\begin{document}
\maketitle
\begin{abstract}

This paper presents a multi-stage reranking system for repository-level code search, which leverages the vastly available commit histories of large open-source repositories to aid in bug fixing. We define the task of repository-level code search as retrieving the set of files from the current state of a code repository that are most relevant to addressing a user's question or bug. The proposed approach combines BM25-based retrieval over commit messages with neural reranking using CodeBERT to identify the most pertinent files. By learning patterns from diverse repositories and their commit histories, the system can surface relevant files for the task at hand. The system leverages both commit messages and source code for relevance matching, and is evaluated in both normal and oracle settings. Experiments on a new dataset created from 7 popular open-source repositories demonstrate substantial improvements of up to 80\% in MAP, MRR and P@1 over the BM25 baseline, across a diverse set of queries, demonstrating the effectiveness this approach. We hope this work aids LLM agents as a tool for better code search and understanding. Our code and results obtained are publicly available\footnote{\texttt{https://github.com/Siddharth-Gandhi/ds}}.

\end{abstract}

\section{Introduction}
\label{intro}
The rise of Large Language Models (LLMs) like GPT-4 \cite{gpt4} has revolutionized software development, enabling powerful code generation and assistance capabilities. Tools like GitHub Copilot, with over 1 million paying users \cite{copilot_demo}, demonstrate the commercial success and developer adoption of these technologies.

However, LLMs still struggle with complex, real-world coding tasks that require understanding code at the repository level, such as bug fixing, which often involves reasoning about function interactions across multiple files and identifying subtle mistakes within large codebases. Even as LLM context windows expand to 100K tokens \cite{anthropic100k} and more \cite{munkhdalai2024leave}, empirical evidence suggests that response quality degrades while the risk of hallucination and cost increases. Research and industry findings \cite{resesarchlongcontext} \cite{pinecone} indicate that a concise, high-quality retrieved context outperforms a longer, lower-quality one. This highlights the potential for well-tuned Information Retrieval (IR) methods like BM25, combined with robust reranking, to greatly narrow the search scope and provide better context to LLMs.

Until recently, the majority of code search and generation research has focused on function or snippet-level queries in isolated contexts, largely ignoring the intricacies of repository-level search. However, the emergence of SWE-Bench \cite{swebench} has provided a standardized benchmark for assessing the capability of LLMs in addressing GitHub issues from prominent open-source repositories. State-of-the-art approaches on SWE-Bench, including SWE-Agent \cite{yang2024sweagent} and Retrieval-Augmented Generation (RAG) methods with Claude 3 Opus \cite{anthropic2024claude3} or GPT-4, have demonstrated promising results by employing basic file search techniques using error messages and stack traces to identify buggy files. Nevertheless, these methods do not fully capitalize on the rich repository-level context available, such as the commit history that reflects code evolution and the underlying rationale behind changes.

\begin{figure*}[htpb]
    \centering
    \includegraphics[width=\linewidth]{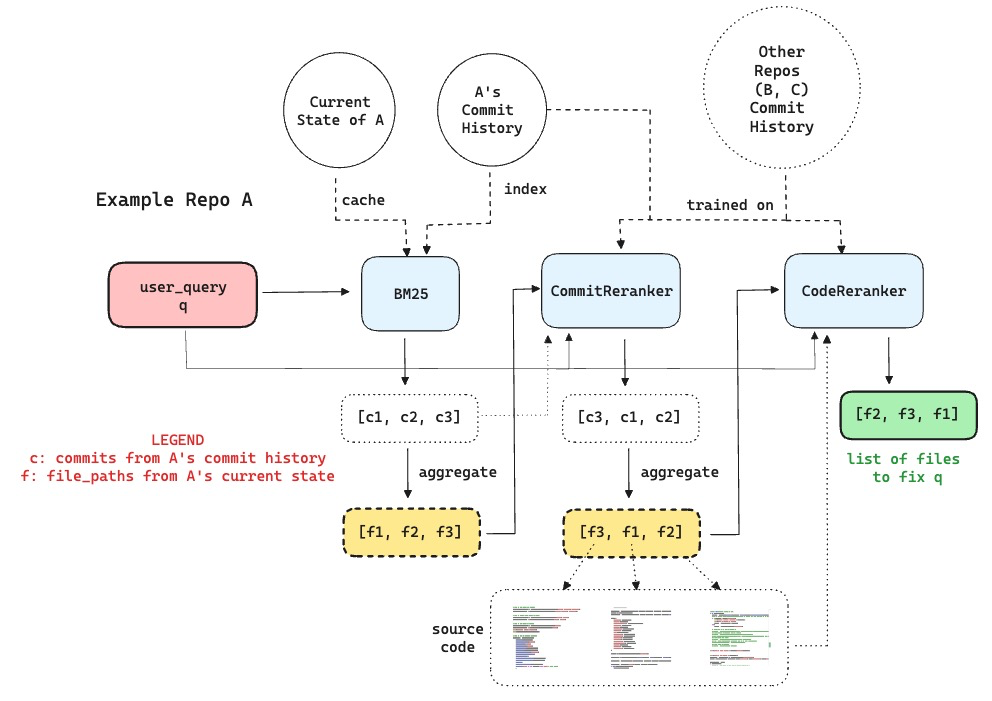}
    \caption{An overview of our system}
    \label{fig:arch}
\end{figure*}

\paragraph{Task Definition} We define the task of repository-level code search as - given a user query $q$, the goal is to retrieve the set of files $\mathcal{F}$ from the \textit{current} state of a code repository that are most relevant to addressing the question or bug described in $q$.
While the search is performed only over the current repository snapshot, the proposed retrieval system can be trained on a diverse collection of code repositories and their associated commit histories.
This enables the model to learn patterns of how similar issues have been resolved across different codebases over time, potentially enhancing its ability to surface pertinent files for the task at hand.

In this paper, we propose a multi-stage ranking system to address the task of repository-level code search, leveraging both commit messages and source code as relevance indicators. Figure \ref{fig:arch} outlines the architecture of our system.

Given a user query at a specific state of the repository, our system outputs a highly relevant ranked list of files most likely to address the query.
The proposed approach consists of three main components:

\begin{enumerate}
    \item A BM25-based system to search over previous commits with similar messages in the repository and identify the files they modified, narrowing down the search scope from the order of 10,000 to approximately 1,000 files.
    \item A BERT-based CommitReranker that reranks the list obtained from the BM25 system, taking into account the semantic understanding of the commit messages. This step further reduces the number of candidate files from the order of 1,000 to hundreds.
    \item A BERT-based CodeReranker that examines the source code of the candidate files and reranks them to achieve maximum precision, ultimately reducing the list from hundreds to tens of files.
\end{enumerate}

 The ultimate goal is to produce a concise set of highly pertinent files that can serve as enriched file context for a powerful language model, such as GPT-4 or CodeLlama \cite{codellama}. This enhanced context can then be utilized to suggest appropriate code modifications that effectively address the specified issues. We hypothesize that this strategy will result in a substantial improvement in the quality and relevance of the generated code.


\section{Related Works}
\label{lit_review}
\subsection{Advances in Information Retrieval}
We do not find any papers that leverage commit messages for the task of code search. However, in general natural-language document search we note many advances. Classic lexical methods, such as BM25 \cite{bm25}, have stood the test of time being computationally efficient but suffer from vocabulary mismatch issues, as they rely on exact term matching and fail to capture semantic similarities between queries and documents. To address this limitation, neural ranking models, including BERT-based approaches like BERT-RR \cite{bertrr} and Dense-RR \cite{denserr}, have been proposed. These models leverage the pre-trained contextual representations of BERT \cite{devlin2018bert} to better understand the semantic relevance between queries and documents, leading to significantly improved retrieval precision compared to traditional lexical methods.

\subsection{Neural Code Models}

Recent advancements in pre-training techniques have significantly improved the performance of language models on code-related tasks. CodeBERT \cite{codebert} employs bimodal pre-training using Masked Language Modeling (MLM) and Replaced Token Detection (RTD) objectives on a dataset of natural language (NL) and programming language (PL) pairs. This approach enables CodeBERT to learn the relationships between NL and PL, leading to improvements in tasks like natural language code search and code documentation generation.

GraphCodeBERT \cite{graphcodebert} extends this idea by incorporating the inherent structure of code into its pre-training, constructing a graph representation of code snippets to capture the relationships between variables, functions, and other code elements. More recently, generative LLMs like CodeT5 \cite{codet5}, StarCoder \cite{starcoder}, and CodeLlama have emerged, specializing in understanding and generating code by leveraging semantic information present in code, such as developer-assigned identifiers and comments. Since we only focus on retrieval and not generation, we will use CodeBERT as our neural retrieval model.

\subsection{Neural Code Search Benchmarks}
Popular code search benchmarks like CodeSearchNet \cite{codesearchnet} and CodeXGLUE \cite{codexglue} focus on retrieving relevant code snippets for natural language queries from a large collection of functions across multiple programming languages. Models such as CodeBERT, GraphCodeBERT, and DenseCode \cite{densecode} tackle this task by encoding both queries and code, computing relevance scores using the \texttt{[CLS]} token or dense representations. However, these benchmarks do not adequately address the more complex task of repository-level code file search, which requires understanding the interconnections and dependencies among various files within a larger codebase, considering the context of changes over time. While benchmarks like SWEBench \cite{swebench}, which consists of 2,294 real-world software engineering problems drawn from GitHub issues and corresponding pull requests across 12 popular Python repositories, propose repository-wide tasks such as editing the codebase to address a given issue, they focus on code completion and only consider bugs involving one file.

\subsection{Repo-Level Code Models and LLM Agents}
Prior works in repository-level code completion have focused on leveraging pre-trained code language models and incorporating project-level context to improve performance. These approaches aim to continue writing unfinished code by considering the broader context of the repository, which is essential in modern modular software development. RepoCoder \cite{repocoder} proposes a framework that combines a similarity-based retriever (over sliding windows of code) with a pre-trained code language model to generate code at line level. CCFinder \cite{ccfinder} integrates cross-file context into pre-trained code language models by building a project context graph based on \texttt{import} statements, and does joint learning of in-file and cross-file context. 

More recently, on SWE-Bench \cite{swebench}, the performance of popular LLMs like GPT-4 and Claude Opus remains limited, with only 1-4\% of tasks solved using Retrieval Augmented Generation (RAG) \cite{rag}. RAG employs BM25 on code files and passes as much context as available within the LLM's context window (max 32K). Stronger LLM-based agents, such as the commercial Devin and open-source agents like Open Devin and SWE-Agent \cite{yang2024sweagent}, have reportedly increased the success rate to 12-15\% on SWE-Bench tasks. However, the internal workings of these agents have not been publicly disclosed.

Interestingly, SWE-Bench notes that even under oracle settings, where the LLM is provided with the relevant file, GPT-4 could only solve 5\% of issues, compared to 1\% under normal conditions. While this suggests that the retriever may not be the primary bottleneck, we hypothesize that vanilla LLMs do not yet fully exploit the iterative nature and feedback loop available in software development (e.g., bugs, error messages). Consequently, we believe that LLM-based agents could benefit from improved retrievers to quickly and reliably identify a small set of relevant files (5-10) to iterate on for a given user bug, potentially leading to more effective bug resolution. Importantly, none of these works have looked into leveraging past commit histories as a mapping to be learnt to solve current user bugs. This is the area we investigate in this work.


\begin{table*}[ht!]
\footnotesize
\ttfamily
\centering
\begin{tabular}{|l|l|}
\hline
\textbf{Column}             & \textbf{Sample} \\ \hline
owner                       & facebook        \\ \hline
repo\_name                  & react        \\ \hline
commit\_date                & 1575406296         \\ \hline
commit\_id                  & f523b2e0d369e3f42938b56784f9ce1990838753        \\ \hline
commit\_message             & Use fewer global variables in Hooks (\#17480)...        \\ \hline
file\_path                  & packages/react-reconciler/src/ReactFiberThrow.js        \\ \hline
previous\_commit\_id        & d75323f65d0f263dd4b0c15cebe987cccf822783        \\ \hline
previous\_file\_content     & @hello ... [content truncated]        \\ \hline
cur\_file\_content          & @world ... [content truncated]        \\ \hline
diff                        & @@ -195,6 +195,18 @@ function throwException(... [content truncated]        \\ \hline
status                      & modified        \\ \hline
is\_merge                   & False          \\ \hline
file\_extension             & js        \\ \hline
\end{tabular}
\caption{Format of data stored in parquet file. \texttt{previous\_*} are set when \texttt{status} is not \texttt{added}, \texttt{is\_merge} is \texttt{True} when \texttt{commit\_id} has $>2$ parents (merge of 2 branches), \texttt{date} is in UNIX format (UTC).}
\label{tab:df_format}
\end{table*}

\begin{table*}[htbp]
\centering
\scriptsize
\ttfamily
\resizebox{\textwidth}{!}{
\begin{tabular}{|c|c|c|c|c|c|c|}
\hline
\textbf{Repository} & \textbf{Total Commits} & \textbf{Total Files Edited} & \multicolumn{2}{c|}{\textbf{Edit Freq Per File}} & \multicolumn{2}{c|}{\textbf{Files Edited Per Commit}} \\ \cline{4-7} 
                    &                        &                             & \textbf{\hspace*{1.25mm} Mean \hspace*{1.25mm}} & \textbf{Median} & \textbf{\hspace*{1.5mm} Mean \hspace*{1.5mm}} & \textbf{Median} \\ \hline

\ttfamily{facebook\_react}    & 11609           & 73765                & 10.1         & 4              & 6.4               & 2       \\ \hline
\ttfamily{angular\_angular}                      & 19464           & 151904               & 7.1          & 3              & 7.8               & 2       \\ \hline
\ttfamily{apache\_spark}      & 33679           & 188006               & 13.3         & 4              & 5.6               & 2       \\ \hline
\ttfamily{apache\_kafka}      & 10445           & 75655                & 9.9          & 4              & 7.2               & 2       \\ \hline
\ttfamily{django\_django}     & 21991           & 81252                & 18.3         & 7              & 3.7               & 2       \\ \hline
\ttfamily{julialang\_julia}                      & 46778           & 182112               & 41.7         & 8              & 3.9               & 2       \\ \hline
\ttfamily{redis\_redis}       & 11077           & 29533                & 15.9         & 2              & 2.7               & 1

\\ \hline\multicolumn{5}{|c|}{} \\[-1em]  
\hline  
\textbf{Average}            & 21006.1         & 111746.7             & 16.6        & 4.5            & 5.3              & 1.8         \\ \hline       \end{tabular}}
\caption{Overall Statistics Per Repository}
\label{app_tab:overall_repo_stats}
\end{table*}

\section{Proposed System}
From Figure \ref{fig:arch}, our system aims to take a user bug as input and outputs a list of potential files relevant to the query. At the repository level, we retrieve files from the current state of the repository and ground it in past commit messages and source code. We also incorporate cross-repository training for our rerankers. We discuss specifics below.

\subsection{Dataset}
At the time of this research, there were no publicly available datasets specifically designed for repository-level code-file search using commit histories. To address this gap, we create a new dataset from scratch, guided by the following objectives:
\begin{itemize}
\item Encompassing a diverse range of programming languages and software projects, ensuring variability in project size, complexity, and domains.
\item Incorporating comprehensive commit histories to facilitate an understanding of the codebase's evolution and the context surrounding changes.
\item Maintaining all versions of each file, along with their corresponding commit IDs, messages, diffs, and status (adhering to the GitHub API convention: \texttt{modified}, \texttt{added}, \texttt{deleted}, or \texttt{renamed}).
\end{itemize}

To efficiently collect this data, we clone repositories locally and employ GitPython\footnote{\scriptsize \url{https://github.com/gitpython-developers/GitPython}}, a library offering optimized and well-tested abstractions for Git commands, to scrape commits effectively. It took approximately 2 full days on 4 SSD-compute nodes to finish the scraping with approximately 150 GB of data (including BM25 index) for 90 popular open source repositories\footnote{\scriptsize List available at \url{https://github.com/Siddharth-Gandhi/ds/blob/boston/misc/repo_info.txt}}. Data for each repository is stored as parquet files, because of its efficient compression, fast reads and cross-platform compatibility. A sample row is shown in Table \ref{tab:df_format}.

Due to compute limitations in subsequent experiments, we currently focus on 7 repositories. Table \ref{app_tab:overall_repo_stats} presents relevant statistics for each repository, including the average number of files edited per commit (to assess commit granularity) and the frequency of edits for individual files throughout the commit history (to identify potentially 'easy' files that are modified across many commits). The median edit frequencies suggests that most edited files are still only in $<2\%$ of commits on average. Thus a Zipfian edit distribution is not likely, and metrics inflation should not be a problem.


\begin{table*}[h]
\centering
\begin{tabular}{@{}p{0.45\textwidth}p{0.45\textwidth}@{}}

\hline

\textbf{\small\ttfamily Original Commit Message} & \textbf{\small\ttfamily GPT-4 Transformed Short Query} 
\\ \hline

\tiny\ttfamily
[facebook\_react] Resolve default onRecoverableError at root init (\#23264) Minor follow up to initial onRecoverableError PR. When onRecoverableError is not provided to `createRoot`, the renderer falls back to a default implementation. Originally I implemented this with a host config method, but what we can do instead is pass the default implementation the root constructor as if it were a user provided one. &
\tiny\ttfamily
`createRoot` doesn't provide a default implementation for `onRecoverableError`, causing the renderer to fail when `onRecoverableError` is not specified.
\\
\tiny\ttfamily
[facebook\_react] Allow the user to opt out of seeing "The above error..." addendum (\#13384) * Remove e.suppressReactErrorLogging check before last resort throw It's unnecessary here. It was here because this method called console.error(). But we now rethrow with a clean stack, and that's worth doing regardless of whether the logging is silenced. * Don't print error addendum if 'error' event got preventDefault() * Add fixtures * Use an expando property instead of a WeakSet * Make it a bit less fragile * Clarify comments &
\tiny\ttfamily
When an 'error' event is preventDefault(), unnecessary error addendum prints are still occurring.

\\ \hline
\end{tabular}
\caption{Examples of transformed queries for two sample commit messages from {\small\ttfamily facebook\_react}.}
\label{tab:transformed_query_examples}
\end{table*}

    \subsection{Evaluation}
    \label{sec:eval}
    At the time of starting this research, SWE-bench was not available publicly. Further, SWE-bench only tests the capability to retrieve one file for a given query, however we envision repo-level code search to retrieve multiple files if necessary. Thus we created our own evaluation data from the commit messages we have available.
    
    \paragraph{GPT-Modified (GPT-M) Test Query Set} To create a realistic evaluation setup, we select 100 commits from each repository as test-commits, which are excluded from the training data for all models. For each test-commit, we feed the commit message to GPT-4 and prompt it to generate the most probable problem description in the style of a GitHub issue, using 1-2 sentences. Importantly, we ask GPT-4 to mask the solution presented in the commit message to ensure that the generated issue does not contain information about the fix. Samples are available in Table \ref{tab:transformed_query_examples} and the complete prompt with additional samples is in Appendix \ref{app:gpt4_prompt}. 
    
    The resulting issue descriptions are typically 30-40 tokens, mimicking real-world scenarios where developers often describe issues concisely without extensive knowledge of the underlying problem or its solution. We also experimented with GPT-3.5 Turbo but found that it often leaked solution details into the modified query, undermining the purpose of this masking step.
    
    \paragraph{Metrics} For this task, we use usual IR metrics to evaluate our system: Mean Average Precision (MAP), Precision at $k$ (P@{1,5,10,20}) Mean Reciprocal Rank (MRR), Recall at $k$ (R@{1,10,20,100,500,1000}). 
    
    
    \paragraph{Relevance matching} For a test query $q_t$, which is a GPT-Modified version of a test commit $c_t$ with a list of actually modified files $F_t$, we consider a file $F_r[i]$ from our system's retrieved list of files $F_r$ to be relevant if it is present in $F_t$. Here, a file refers to its file path. However, this approach has a limitation: not all files in $F_t$ may be relevant to $q_t$, as different GitHub repositories vary in their commit granularity and the level of detail in commit messages, ranging from many small commits with precise changes to a single large commit with multiple changes not fully described in the message. Table \ref{tab:test_set_stats} shows the number of relevant files per test query for each repository.
    
\begin{table}[ht!]
\centering
\scriptsize
\ttfamily

\begin{tabular}{|l|c|c|}
\hline
\textbf{Repository} & \multicolumn{2}{c|}{\textbf{Files Edited Per Query}} \\ \cline{2-3} 
                     & \textbf{\hspace*{1.5mm} Mean \hspace*{1.5mm}} & \textbf{Median} \\ \hline
\ttfamily{facebook\_react}            & 5.1                & 4                     \\ \hline
\ttfamily{angular\_angular}           & 3.5                & 2                     \\ \hline
\ttfamily{apache\_spark}              & 3.9                & 3                     \\ \hline
\ttfamily{apache\_kafka}              & 5.4                & 3                     \\ \hline
\ttfamily{django\_django}             & 2                  & 2                     \\ \hline
\ttfamily{julialang\_julia}          & 2.4                & 2                     \\ \hline
\ttfamily{redis\_redis}               & 2.4                & 2                     \\ \hline
\multicolumn{3}{|c|}{} \\[-1em]  
\hline  
\textbf{Average}           & 3.5               & 2.8                     \\ \hline
\end{tabular}
\caption{Test Query Set Statistics Per Repository}
\label{tab:test_set_stats}
\end{table}

\subsection{BM25 Initial Ranker}
\label{sec:bm25}
To identify relevant files for a given user query, we first employ BM25 to calculate similarity scores between the query and commit messages (as shown in Table \ref{tab:df_format}). We select BM25 \cite{bm25} as a baseline and initial ranker for further rerankers because of its efficiency and high recall over a variety of datasets. We use Pyserini's \cite{Lin_etal_SIGIR2021_Pyserini} implementation of BM25 in Python.

The objective of matching commits is to determine whether a similar issue has been encountered before in the repository, either exactly or approximately. BM25 retrieves relevant commits, but each commit can modify multiple files simultaneously. However, our ultimate goal is to retrieve a set of files, not commits. To achieve this, we aggregate the BM25 scores across the files, considering the scores of all commits that have modified each file. This aggregation process allows us to identify the most relevant files based on their association with commits that are similar to the given query.

\paragraph{File Based Aggregation} Say for a given query $q$, BM25 retrieves commits $[c_1, c_2, c_3]$ where $c_1$ had edited files $[f_1, f_2, f_3]$ and received a BM25 score 5, $c2$ had files $[f_2, f_5, f_1, f_6]$ with score 3, and $c3$ had $[f_4, f_3, f_2, f_7, f_8]$ with score 1. Then we can aggregate scores across files as $[f_2: 9, f_1: 8, f_3: 6...]$ using various aggregation strategies like \mono{sump} (shown), \mono{maxp} or \mono{avgp}. In our preliminary experiments, we found that \mono{maxp} performed the best (in R@1000), so we use that for the remaining experiments.

\paragraph{Time Masking} As our evaluation is based on test commits, which represent past repository states, and BM25 searches over all available commits, it is essential to apply masking to filter out retrieved future commits from the search results. This prevents information leakage from future commits when evaluating historical repository states, maintaining the integrity of the evaluation and eliminating any unfair advantage. It is important to note that in a real-time deployment scenario, masking would not be necessary since there would be no future versions of the files available at the time of the search.

\paragraph{Tokenization} We use TikToken \cite{tiktoken}, a fast Byte Pair Encoding (BPE) tokenizer trained on both code and natural language, to tokenize the commit messages when creating search index for each repository. This approach is more suitable for queries like GitHub issues or user bugs, which often contain programming snippets, compared to Pyserini's built-in lexical tokenizers designed for natural language. Table \ref{app_tab:repo_token_stats} shows token statistics over commit messages and GPT-Modified query set.

\begin{table}
\centering
\scriptsize
\ttfamily

\begin{tabular}{|l|c|c|c|}
\hline
\multicolumn{1}{|c|}{\textbf{Repository}}  & \textbf{\begin{tabular}[c]{@{}c@{}}Commit \\  Messages \\ Avg Tokens\end{tabular}} & \textbf{\begin{tabular}[c]{@{}l@{}}GPT-M \\  Queries \\ Avg Tokens\end{tabular}} & \textbf{\begin{tabular}[c]{@{}l@{}}File Path \\ Avg Tokens\end{tabular}} \\ \hline
\ttfamily{facebook\_react}                   & 242.1   & 36.1    & 20.9        \\ \hline
\ttfamily{angular\_angular}                  & 197.1      & 36.7              & 20.3      \\ \hline
\ttfamily{apache\_spark}                     & 819.9   & 39.6     & 30.3    \\ \hline
\ttfamily{apache\_kafka}                     & 254.9   & 37.1     & 28.1   \\ \hline
\ttfamily{django\_django}                    & 59.9     & 27.3               & 14        \\ \hline
\ttfamily{julialang\_julia}                  & 124.3    & 32.2    & 9     \\ \hline
\ttfamily{redis\_redis}                      & 236.1    & 37.5    & 8.1    \\ \hline
\multicolumn{4}{|c|}{} \\[-1em]  
\hline
\textbf{Average}    & 276.3                        & 35.1     & 18.7          \\ \hline
\end{tabular}
\caption{Token Statistics for training sets per repository} 
\label{app_tab:repo_token_stats}
\end{table}

\begin{figure}[htpb]
    \centering
    \includegraphics[width=\linewidth]{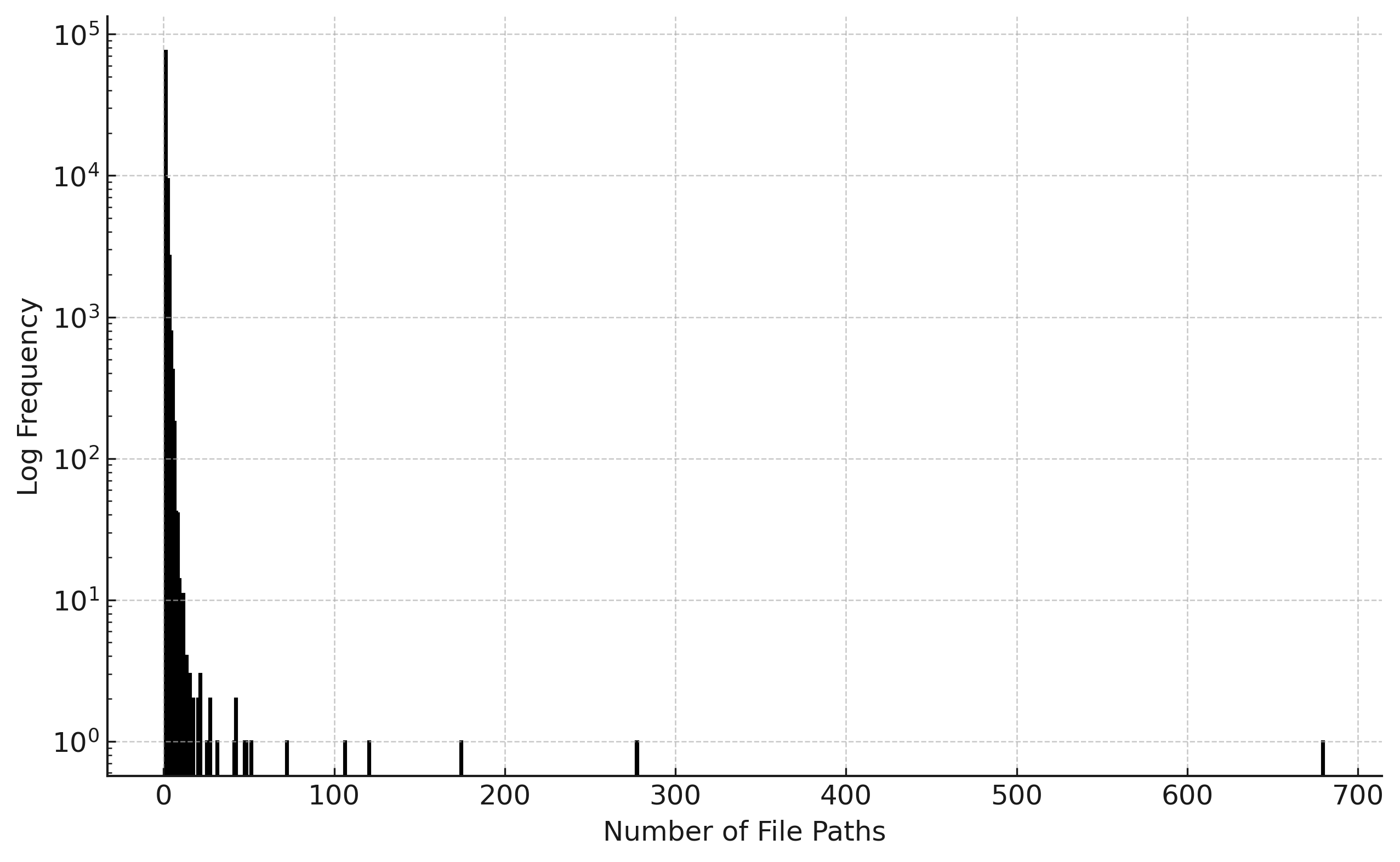}
    \caption{Histogram of Number of files per FID across 7 repositories}
    \label{fig:fid_hist}
\end{figure}

\paragraph{File ID (FID) Mapping and Filtering for Current Repository State} File paths in a repository can change over time due to deletions, renaming, or branching. To address this issue and ensure that the retrieved files exist in the current repository state, we employ a two-stage approach. First, we parse all commits in the repository and create FID objects, where the FID of a file $f$ represents all possible file paths that the file has ever had. These mappings are stored as caches from FID to file path and vice versa. Figure \ref{fig:fid_hist} shows the distribution of the number of file paths mapped to each File ID (FID). The majority of FIDs map to 1-5 files, but there are a few outliers where a single FID maps to hundreds of files, such as \texttt{init} files or multi-language documentation. Second, after retrieving relevant commits using BM25, we filter out FIDs that do not exist in the current repository state, which has been previously cached. This process guarantees that the output consists of valid files present in the repository's current state and reduces the retrieved BM25 file list by half.

It is important to note that our approach relies on Git's internal tracking of file renames, which is based on heuristics and may not be entirely accurate. Consequently, there are cases where a single FID may map to two different files at the same repository state, which should never occur. In such instances, we currently select one of the file paths. The algorithm for computing file paths is linked in Appendix \ref{app:fid}, along with a discussion on the inaccuracies in Git's tracking.

After this initial retrieval, including retrieving commits, aggregating scores, masking future commits, and filtering invalid file paths, BM25 ultimately provides us with a ranked list of files that it considers highly likely to address the user query $q$ based on the similarity of the query to past commit messages. This reduces the search space from 10,000+ files to about 1000 files.

\subsection{CommitReranker: Better Commit Message Matching}
\label{sec:bertrr}

While BM25 effectively retrieves relevant commits, it relies on exact token matches and may not capture the semantic meaning and context of each message, leading to vocabulary mismatch issues. To address this and improve the rankings, particularly the Recall@100 (R@100), we employ a BERT-based reranker \cite{bertrr} on the commit messages to rerank the top $k$ files. BERT first embed the query and commit messages (passage) together, separated by the \texttt{[SEP]} token. Relevance scores are obtained by passing the \texttt{[CLS]} token representation through a linear layer, capturing the semantic similarity between the query and the passage.

Our ultimate goal is to incorporate the source code itself into our ranking pipeline. However, computing this for over 1,000 files might be inefficient. By using a BERT reranker to mitigate vocabulary mismatches and improve the predicted scores of (Query, Retrieved Commit Message) pairs, we aim to enhance the R@100 and obtain better rankings of file paths through aggregation.


\paragraph{One File, Multiple Commit Mapping}
For a given (query, file) pair, the CommitReranker considers all commits $C=[c_1, c_2, c_3]$ in which the file was modified. We compute a relevance score for each (query, commit message) pair and aggregate the individual commit scores to obtain a single score for the file. We experimented with different aggregation methods and found that taking the maximum score (\texttt{maxp}) across all commits yielded the best results, aligning with the intuition that only one of the file modifications throughout the commits might be relevant to the current issue, while others may address different concerns. 

\paragraph{Supervised Contrastive Finetuning}
To adapt CodeBERT for the specific task of commit message reranking, we finetune it using supervised contrastive training on (query, passage, label) triplets constructed from past training data. The training data is obtained as follows:

\begin{itemize}
\item \textbf{Query:} A query $q_{train}$ from a set of training queries $Q_{train}$, which are GPT-modified versions of a fixed number of train commits (similar to Section \ref{sec:eval}). Each query has a ground truth list of edited files $F_{truth}$. We also found that training on just the commit messages to be equivalent. 

\item \textbf{Passages:} For a given query $q_{train}$, we retrieve pre-file aggregation BM25 results $R_{bm25} = [c_1, c_2,...]$. This step effectively finds commit messages similar to the user query. Note that each commit $c_i$ can have multiple associated files $F_{c_i}$.
\item \textbf{Label:} The task now becomes determining if a given query and retrieved commit message are related to a similar issue. As this is a non-trivial task, we rely on heuristics. One approach is to take the intersection between $F_{truth}$ and $F_{c_i}$ and consider it positive (label = 1) if it is non-zero, otherwise negative (label = 0). Another approach is to compare the diffs $d_{truth}$ and $d_{c_i}$ line by line (expanded in Section \ref{sec:coderr}) to assign labels. However, in our testing, we did not find a significant difference between these approaches.
\end{itemize}

\paragraph{Loss Function} The model is trained using the Mean Squared Error (MSE) loss between the predicted score and the label, with the objective of assigning higher scores to (passage) commits that have edited files similar to those of the (query) commits. This training process aims to capture the model's understanding of the underlying motivation behind the code changes. We also experimented with using cross-entropy loss for finetuning; however, we found that it yielded suboptimal results compared to MSE.

Thus, for a given (query, commit message) pair, CommitReranker predicts a score between 0 and 1, indicating the similarity between the query and the commit message.

\subsection{CodeReranker: Integrating Code File Relevance}
\label{sec:coderr}
Next, we develop a model that incorporates relevance indicators from the code files themselves. We finetune another \texttt{microsoft/codebert-base} model using the same loss as in Section \ref{sec:bertrr}, embedding the query and passage separated by the \texttt{[SEP]} token to obtain relevance scores. Training is also supervised contrastive with triplets constructed as follows for a set of training queries $Q_{train}$: 
\begin{itemize}
\item \textbf{Query:} A query $q_{train} \in Q_{train}$, remains the same as in Section \ref{sec:bertrr}.
\item \textbf{Passages (Code Patches):} For $q_{train}$, we retrieve file-aggregated BM25 results $R_{bm25} = [f_1, f_2,..]$, yielding a list of candidate source code files for the query. Since CodeBERT has a sequence length of 512 tokens and files in \texttt{facebook\_react} have an average of 27,000 tokens (~75 patches for a 350-token patch length), we need to split the code files into smaller code patches. We experiment with three splitting strategies:
\begin{enumerate}
\item Random splitting of tokens until a patch length of 350 is reached
\item Function parsing, converting source code into a syntax tree with functions and classes, and considering each function as a code patch
\item Line-wise splitting, using a sliding window over lines and starting the next window once 350 tokens are reached in the current window
\end{enumerate}
We found that line-wise splitting (c) performed best, followed by function parsing (b) and random splitting (a). Function parsing incurs additional overhead and does not guarantee that the resulting code patches fit within 350 tokens.
\item \textbf{Label:} For each query $q_{train}$ with a list of ground truth files $F_{truth}$, we have granular diffs $d_f$ for each file $f \in F_{truth}$. Similarly, for each retrieved file, we have diffs from the candidate commit obtained through previous ranker's \texttt{maxp} aggregation. We perform a line-by-line comparison of the diffs, considering a code patch as positive (label = 1) if there is a non-zero intersection (after removing trivial lines like brackets and newlines), and negative (label = 0) otherwise.
\end{itemize}

Thus, for a given (query, file) pair, the CodeReranker splits the file into code patches and computes a relevance score for each (query, code patch) pair. The final score for the file is obtained by aggregating the individual code patch scores. We experimented with different aggregation methods and found that taking \texttt{maxp} aggregation across all code patches yielded the best results. This aligns with the intuition that only a few specific sections of a file might be relevant for addressing a particular query, rather than the entire file.

\subsection{Experimental Setup}
\label{sec:exp_setup}

\paragraph{Research Questions}
In our experiments, we aim to address the following research questions:
\begin{enumerate}
\item Is leveraging past commit messages a viable approach for repository-level code search? Does this method yield reasonable results?
\item How does the reranking depth influence the performance of the rerankers? We investigate the impact of varying reranking depths, specifically 100, 250, 500, and 1,000.
\item How do the CommitReranker and CodeReranker perform individually under optimal conditions (i.e., in an oracle setting)?
\item How does the complete pipeline, combining BM25, CommitReranker, and CodeReranker, perform on diverse sets of test queries across different repositories?
\end{enumerate}

\paragraph{General} We experiment with 7 different repositories (Table \ref{app_tab:overall_repo_stats}), where each has their own BM25 commit index, 500 GPT-M train queries and 100 GPT-M test queries. 

\paragraph{Cross-Repository Combined Training of Rerankers}
We discovered that training the CommitReranker and CodeReranker on combined data from multiple repositories (the 7 repositories used in our experiments and 3 additional ones) with a total of 5,000 training queries yielded better results compared to training on a single repository, even with a larger number of train commits. This finding suggests that the diversity of data across repositories is more beneficial for the rerankers' performance than a larger amount of training data from a single repository. The cross-repository training approach exposes the models to a wider variety of commit messages, code changes, and query-file relationships, enabling them to generalize better to unseen queries and repositories.

\paragraph{BM25 (Baseline)} We use $k1=0.9,b=0.4$ in Pyserini BM25 for initial ranking. For file aggregation, we use \mono{maxp} in all our experiments to maximize Recall@100. We retrieve at most $k=1000$ initial files for initial retrieval. \texttt{p50k\_base} model from tiktoken was the tokenizer for the commit message index.


\paragraph{Training Details for CommitReranker and CodeReranker} Both models are trained on an Nvidia Quadro RTX6000 GPU with 25 GB VRAM. We use a batch size of 32 and train the models for 8 epochs with a learning rate of $5*10^{-5}$. We employ the \texttt{maxp} aggregation strategy for both models. Following the triplet data curation process described in Section \ref{sec:bertrr} for the CommitReranker, we obtain 38,240 positive and 38,240 negative (query, commit message, label) triplets. For the CodeReranker, as outlined in Section \ref{sec:coderr}, we have 11,202 positive and 11,202 negative (query, code patch, label) triplets. Both use \texttt{codebert} tokenizer. For each train query, we retrieve BM25 results and take first 10 positive and first 10 negatives for triplet creation.

\paragraph{Oracle Setting} To assess the individual performance of both CommitReranker and CodeReranker under optimal conditions, we create an oracle setting. Instead of having BM25 as the initial retriever, we randomly distribute the correct files across various reranking depths.

\paragraph{Full Pipeline} This involves BM25 with initial retrieval depth of 1000, CommitReranker at rerank depth 1000 and finally CodeReranker at rerank depth 100.

\section{Results \& Discussion}
\label{results}

\begin{figure*}[htpb]
    \centering
    \includegraphics[width=\linewidth]{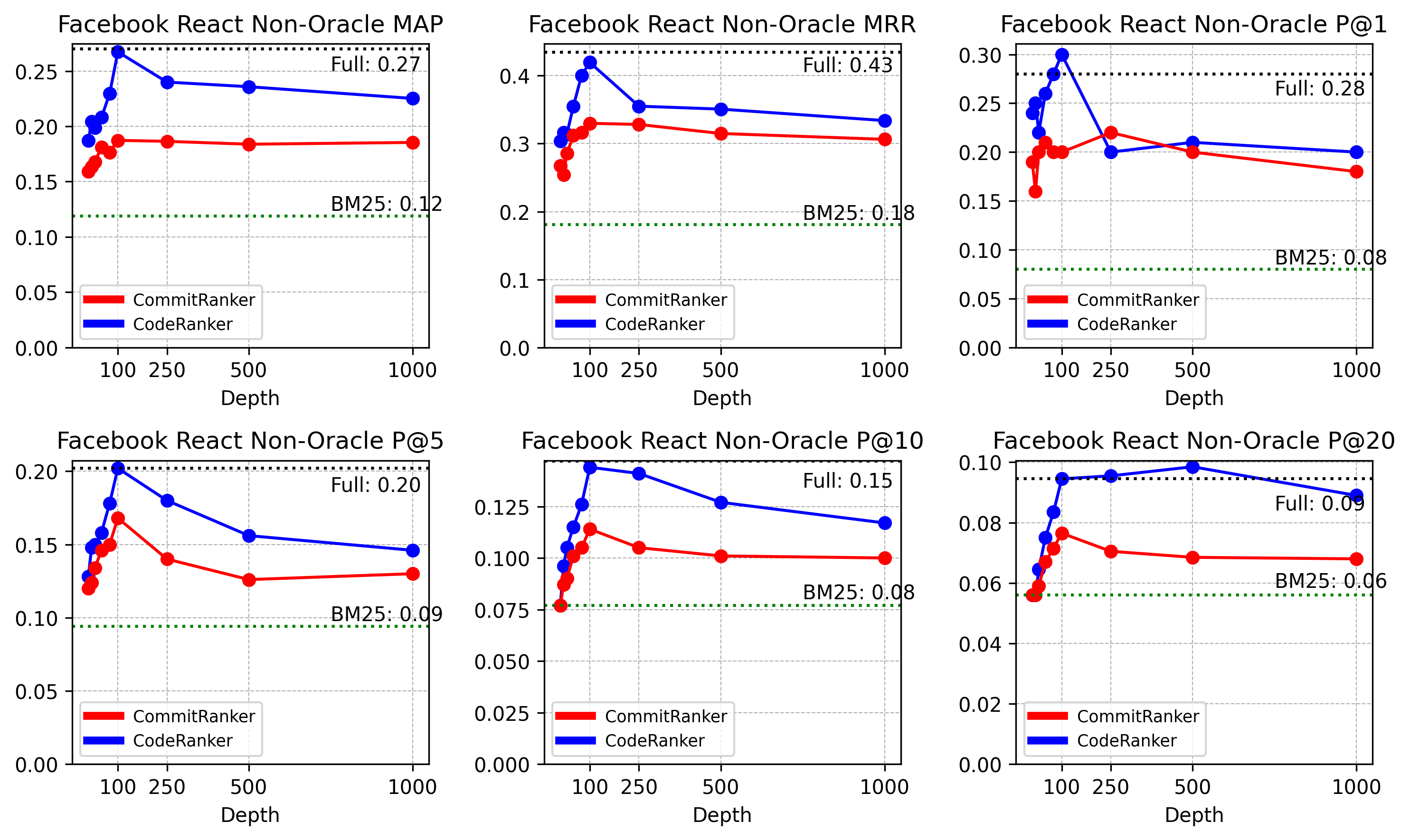}
    \caption{Visualization of Table \ref{tab:fbr_all}a. CodeReranker (blue) is significantly better compared to CommitReranker(red) in all settings, however the Full Pipeline is still surprisingly better.}
    \label{fig:fbr_non_oracle}
\end{figure*}

\begin{figure*}[htpb]
    \centering
    \includegraphics[width=\linewidth]{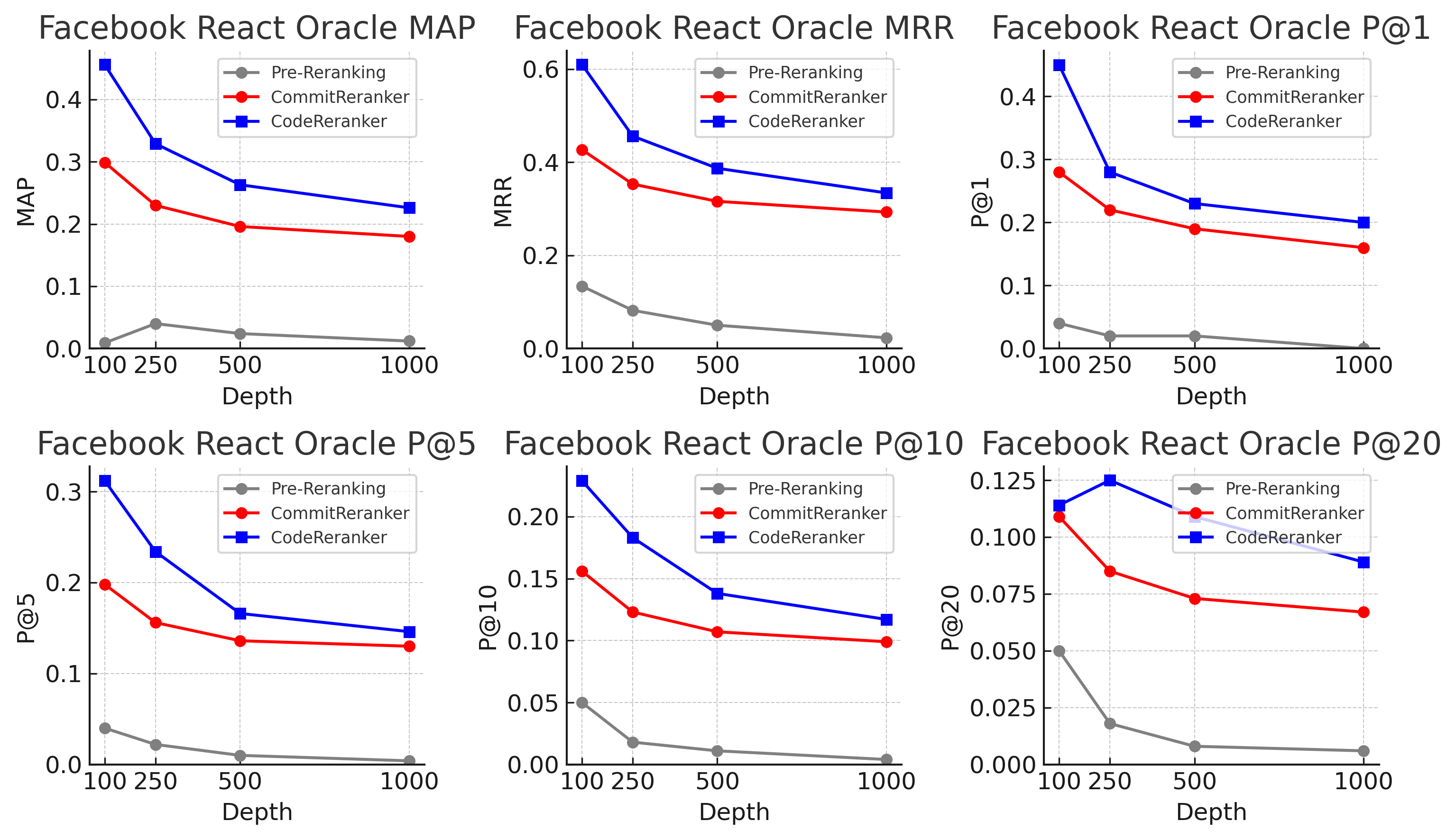}
    \caption{Visualization of Table \ref{tab:fbr_all}b. Notice how much the grey curve is lifted by both red and blue lines. CodeReranker (blue) improves the same pre-ranking (grey) significantly more than CommitReranker (red).}
    \label{fig:fbr_oracle}
\end{figure*}

This section studies results listed in next 3 pages. Table \ref{tab:combined_results} and Figure \ref{fig:res} contain averaged results across 7 repositories. Per repository results are available in Appendix \ref{app:per_repo_results}. Meanwhile, Table \ref{tab:fbr_all} and Figures \ref{fig:fbr_non_oracle}, and \ref{fig:fbr_oracle} look at just one particular repository (\texttt{facebook\_react}) in much more detail in both oracle and normal settings. An important number to keep in mind from Table \ref{tab:test_set_stats} is, that there are approximately 3 files edited per test query. Since the goal is to rank these as high as possible, the most important metrics are P@1 and MRR.

\begin{table}[htbp]
\centering
\ttfamily
\scriptsize
\begin{tabular}{|l|r|r|r|}
\hline
\textbf{Type} & \textbf{Depth} & \textbf{Total Time} & \textbf{Per Query Time} \\
\hline
BM25 & - & 7.5 mins & 4.5 s / query \\
\hline
\multirow{4}{*}{Commit Reranker} & 100 & 1.5 hrs & 1 min / query \\
 & 250 & 3.5 hrs & 2 mins / query \\
 & 500 & 5 hrs & 3 mins / query \\
 & 1000 & 5.5 hrs & 3.5 mins / query \\
\hline
\multirow{4}{*}{Code Reranker} & 100 & 1.5 hrs & 1 min / query \\
 & 250 & 3.5 hrs & 2 mins / query \\
 & 500 & 6 hrs & 3.2 mins / query \\
 & 1000 & 7.5 hrs & 4.5 mins / query \\
\hline
Full & - & 7.5 hrs & 4.5 mins / query \\
\hline
\end{tabular}
\caption{Summary of total and per query time for different methods on \texttt{facebook\_react} for 100 test queries}
\label{tab:query_times}
\end{table}

\paragraph{BM25} demonstrates reasonably good performance while being quite efficient, with Recall being the most important metric since it serves as the initial ranker. The Recall@1000 is nearly perfect, around 0.92, indicating that almost all relevant files are retrieved within the top 1,000 results. Surprisingly, we also observe a decent Mean Reciprocal Rank (MRR) of 0.22 and Precision@1 (P@1) of 0.139, suggesting that by simply performing exact matching against past commit messages, we can retrieve some useful files and potentially resolve user queries. We further investigated BM25 results at depths of 2,000, 5,000, and 10,000; however, the recall tapers off, implying that some files are either very challenging to retrieve (due to being committed with an unrelated message) or suffer from the FID issue mentioned in Section \ref{sec:bm25}.


\begin{table*}[htbp]
\centering
\ttfamily
\scriptsize
\resizebox{\textwidth}{!}{\begin{tabular}{|l|r|r|r|r|r|r|r|r|r|r|r|r|r|}
\hline
 \textbf{Type} & \textbf{Depth} & \textbf{MAP} & \textbf{MRR} & \textbf{P@1} & \textbf{P@5} & \textbf{P@10} & \textbf{P@20} & \textbf{R@1} & \textbf{R@10} & \textbf{R@20} & \textbf{R@100} & \textbf{R@500} & \textbf{R@1000} \\

\hline
\textbf{BM25} & \textbf{} & 0.163 & 0.227 & 0.139 & 0.090 & 0.071 & 0.048 & 0.061 & 0.265 & 0.336 & 0.607 & 0.846 & 0.924 \\
\cline{1-14}
\multirow[t]{4}{*}{\textbf{CommitReranker}} & \textbf{100} & 0.252 & 0.361 & 0.250 & 0.143 & 0.103 & 0.067 & 0.114 & 0.390 & 0.476 & 0.607 & 0.846 & 0.924 \\
\textbf{} & \textbf{250} & 0.249 & 0.348 & 0.240 & 0.134 & 0.096 & 0.067 & 0.112 & 0.368 & 0.493 & 0.704 & 0.846 & 0.924 \\
\textbf{} & \textbf{500} & 0.236 & 0.331 & 0.220 & 0.123 & 0.091 & 0.064 & 0.103 & 0.359 & 0.473 & 0.702 & 0.846 & 0.924 \\
\textbf{} & \textbf{1000} & 0.229 & 0.317 & 0.210 & 0.118 & 0.087 & 0.061 & 0.099 & 0.352 & 0.454 & 0.709 & 0.886 & 0.924 \\
\cline{1-14}
\multirow[t]{4}{*}{\textbf{CodeReranker}} & \textbf{100} & 0.288 & 0.408 & 0.289 & 0.163 & 0.112 & 0.072 & 0.136 & 0.421 & 0.513 & 0.607 & 0.846 & 0.924 \\
\textbf{} & \textbf{250} & 0.275 & 0.389 & 0.259 & 0.156 & 0.115 & 0.076 & 0.121 & 0.438 & 0.539 & 0.723 & 0.846 & 0.924 \\
\textbf{} & \textbf{500} & 0.258 & 0.365 & 0.233 & 0.142 & 0.105 & 0.074 & 0.111 & 0.399 & 0.524 & 0.752 & 0.846 & 0.924 \\
\textbf{} & \textbf{1000} & 0.246 & 0.352 & 0.224 & 0.136 & 0.096 & 0.069 & 0.105 & 0.373 & 0.503 & 0.753 & 0.895 & 0.924 \\
\cline{1-14}
\textbf{Full Pipeline} & \textbf{} & 0.307 & 0.434 & 0.299 & 0.175 & 0.123 & 0.079 & 0.139 & 0.470 & 0.572 & 0.709 & 0.886 & 0.924 \\
\cline{1-14}
\hline
\end{tabular}
}
\caption{Various test-set metrics at varying depths for BM25 Baseline, CommitReranker (red), CodeReranker (blue), and Full Pipeline, averaged across 7 repositories}
\label{tab:combined_results}
\end{table*}

\begin{figure*}[htpb]
    \centering
    \includegraphics[width=\linewidth]{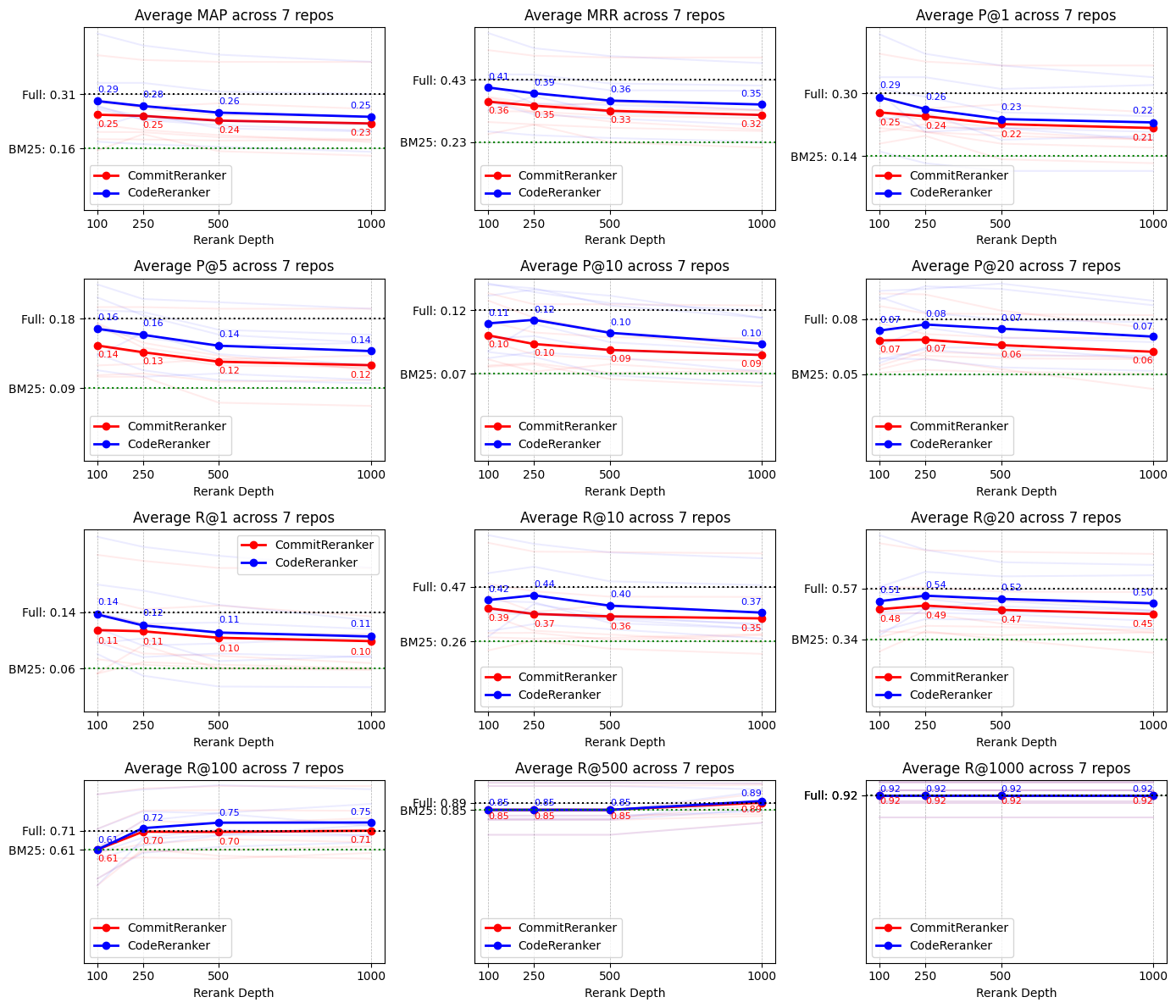}
    \caption{Visualization of Table \ref{tab:combined_results}. Bold lines are averages across 7 repositories, light background lines are individual performance of each constituent repository. Performance does not improve with increasing reranking depths. CodeReranker@100 performs almost to Full Pipeline, probably because of minor R@100 increase with intermediate CommitReranker@1000.}
    \label{fig:res}
\end{figure*}

\begin{table*}
    \centering
    \ttfamily
    \scriptsize
    \subsection*{\ttfamily{[a] Repository: facebook\_react Type: Non-Oracle}}
    \begin{tabular}{|l|r|r|r|r|r|r|r|r|r|r|r|r|r|}
    \hline
    \textbf{Type} & \textbf{Depth} & \textbf{MAP} & \textbf{MRR} & \textbf{P@1} & \textbf{P@5} & \textbf{P@10} & \textbf{P@20} & \textbf{R@1} & \textbf{R@10} & \textbf{R@20} & \textbf{R@100} & \textbf{R@500} & \textbf{R@1000} \\
    \hline
    \textbf{BM25} & \textbf{} & 0.119 & 0.180 & 0.080 & 0.094 & 0.077 & 0.056 & 0.022 & 0.189 & 0.260 & 0.557 & 0.797 & 0.951 \\
    \cline{1-14}
    \multirow[t]{9}{*}{\textbf{CommitReranker}} & \textbf{10} & 0.159 & 0.267 & 0.19 & 0.12 & 0.077 & 0.056 & 0.057 & 0.189 & 0.26 & 0.557 & 0.797 & 0.951 \\
    \textbf{} & \textbf{20} & 0.163 & 0.254 & 0.16 & 0.124 & 0.087 & 0.056 & 0.051 & 0.229 & 0.26 & 0.557 & 0.797 & 0.951 \\
    \textbf{} & \textbf{30} & 0.168 & 0.285 & 0.2 & 0.134 & 0.09 & 0.059 & 0.06 & 0.218 & 0.281 & 0.557 & 0.797 & 0.951 \\
    \textbf{} & \textbf{50} & 0.181 & 0.312 & 0.21 & 0.146 & 0.101 & 0.067 & 0.071 & 0.251 & 0.314 & 0.557 & 0.797 & 0.951 \\
    \textbf{} & \textbf{75} & 0.176 & 0.317 & 0.2 & 0.15 & 0.105 & 0.072 & 0.055 & 0.258 & 0.331 & 0.557 & 0.797 & 0.951 \\
    \textbf{} & \textbf{100} & 0.187 & 0.330 & 0.200 & 0.168 & 0.114 & 0.076 & 0.053 & 0.280 & 0.373 & 0.557 & 0.797 & 0.951 \\
    \textbf{} & \textbf{250} & 0.186 & 0.328 & 0.220 & 0.140 & 0.105 & 0.070 & 0.069 & 0.276 & 0.368 & 0.565 & 0.797 & 0.951 \\
    \textbf{} & \textbf{500} & 0.184 & 0.315 & 0.200 & 0.126 & 0.101 & 0.068 & 0.066 & 0.279 & 0.356 & 0.559 & 0.797 & 0.951 \\
    \textbf{} & \textbf{1000} & 0.185 & 0.306 & 0.180 & 0.130 & 0.100 & 0.068 & 0.061 & 0.293 & 0.369 & 0.590 & 0.874 & 0.951 \\
    
    \cline{1-14}
    \multirow[t]{4}{*}{\textbf{CodeReranker}} & \textbf{10} & 0.187 & 0.304 & 0.24 & 0.128 & 0.077 & 0.056 & 0.083 & 0.189 & 0.26 & 0.557 & 0.797 & 0.951 \\
    \textbf{} & \textbf{20} & 0.204 & 0.317 & 0.25 & 0.148 & 0.096 & 0.056 & 0.086 & 0.243 & 0.26 & 0.557 & 0.797 & 0.951 \\
    \textbf{} & \textbf{30} & 0.199 & 0.314 & 0.22 & 0.15 & 0.105 & 0.064 & 0.073 & 0.264 & 0.299 & 0.557 & 0.797 & 0.951 \\
    \textbf{} & \textbf{50} & 0.208 & 0.355 & 0.26 & 0.158 & 0.115 & 0.075 & 0.081 & 0.301 & 0.355 & 0.557 & 0.797 & 0.951 \\
    \textbf{} & \textbf{75} & 0.23 & 0.401 & 0.28 & 0.178 & 0.126 & 0.084 & 0.081 & 0.323 & 0.392 & 0.557 & 0.797 & 0.951 \\
    \textbf{} & \textbf{100} & 0.267 & 0.420 & 0.300 & 0.202 & 0.144 & 0.094 & 0.097 & 0.382 & 0.465 & 0.557 & 0.797 & 0.951 \\
    \textbf{} & \textbf{250} & 0.240 & 0.355 & 0.200 & 0.180 & 0.141 & 0.096 & 0.076 & 0.374 & 0.477 & 0.630 & 0.797 & 0.951 \\
    \textbf{} & \textbf{500} & 0.236 & 0.351 & 0.210 & 0.156 & 0.127 & 0.098 & 0.081 & 0.352 & 0.514 & 0.683 & 0.797 & 0.951 \\
    \textbf{} & \textbf{1000} & 0.225 & 0.334 & 0.200 & 0.146 & 0.117 & 0.089 & 0.076 & 0.332 & 0.470 & 0.740 & 0.901 & 0.951 \\
    \cline{1-14}
    \textbf{Full Pipeline} & \textbf{} & 0.270 & 0.434 & 0.280 & 0.202 & 0.147 & 0.094 & 0.109 & 0.393 & 0.480 & 0.590 & 0.874 & 0.951 \\
    \cline{1-14}
    \hline
    \end{tabular}

    \subsection*{\ttfamily{[b] Repository: facebook\_react Type: Oracle}}
    \subsubsection*{\scriptsize \ttfamily{Pre-Reranking Metrics (meaning the rankings to be reranked with with Recall@Depth = 1 )}}
    \begin{tabular}{|l|r|r|r|r|r|r|r|r|r|r|r|r|r|}
    \hline
    \textbf{Depth} & \textbf{MAP} & \textbf{MRR} & \textbf{P@1} & \textbf{P@5} & \textbf{P@10} & \textbf{P@20} & \textbf{P@30} \\
    \hline
    \cline{1-8}
    \multirow[t]{4}{*} {\textbf{100}} & 0.009 & 0.134 & 0.04 & 0.04 & 0.05 & 0.05 & 0.05 \\
    \textbf{250} & 0.04 & 0.082 & 0.02 & 0.022 & 0.018 & 0.018 & 0.017 \\
    \textbf{500} & 0.024 & 0.05 & 0.02 & 0.01 & 0.011 & 0.008 & 0.009 \\
    \textbf{1000} & 0.012 & 0.023 & 0.0 & 0.004 & 0.004 & 0.006 & 0.006 \\
    \cline{1-8}
    \cline{1-8}
    \hline
    \end{tabular}

    \subsubsection*{\scriptsize \ttfamily{Post-Reranking Metrics}}
    \begin{tabular}{|l|r|r|r|r|r|r|r|r|r|r|r|r|r|}
    \hline
    \textbf{Type} & \textbf{Depth} & \textbf{MAP} & \textbf{MRR} & \textbf{P@1} & \textbf{P@5} & \textbf{P@10} & \textbf{P@20} & \textbf{P@30} \\
    \hline
    \cline{1-9}
    \multirow[t]{4}{*}{\textbf{CommitReranker}} & \textbf{100} & 0.299 & 0.427 & 0.28 & 0.198 & 0.156 & 0.109 & 0.096 \\
    \textbf{} & \textbf{250} & 0.23 & 0.353 & 0.22 & 0.156 & 0.123 & 0.085 & 0.077 \\
    \textbf{} & \textbf{500} & 0.196 & 0.316 & 0.19 & 0.136 & 0.107 & 0.073 & 0.066 \\
    \textbf{} & \textbf{1000} & 0.18 & 0.293 & 0.16 & 0.13 & 0.099 & 0.067 & 0.057 \\
    \cline{1-9}
    \multirow[t]{4}{*}{\textbf{CodeReranker}} & \textbf{100} & 0.456 & 0.61 & 0.45 & 0.312 & 0.229 & 0.151 & 0.114 \\
    \textbf{} & \textbf{250} & 0.329 & 0.456 & 0.28 & 0.234 & 0.183 & 0.125 & 0.096 \\
    \textbf{} & \textbf{500} & 0.263 & 0.387 & 0.23 & 0.166 & 0.138 & 0.109 & 0.083 \\
    \textbf{} & \textbf{1000} & 0.226 & 0.334 & 0.2 & 0.146 & 0.117 & 0.089 & 0.072 \\
    \cline{1-9}
    \cline{1-9}
    \hline
    \end{tabular}
    \caption{Detailed Results on Facebook React - \textbf{(a) Non-Oracle Results} Normal Rankings at different Rerank Depths Across Different Configurations, \textbf{(b) Oracle Results} Pre-Ranking Metrics (meaning lists that were passed to both CommitReranker and CodeReranker) and Post-Ranking Metrics to have a fair comparison}
    \label{tab:fbr_all}
    \end{table*}

\paragraph{CommitReranker} provides a substantial improvement in all metrics, particularly MRR and P@1, suggesting that there was significant vocabulary mismatch occurring in BM25 matching, which the CommitReranker helps to mitigate. However, as the reranking depth increases, key metrics decline, indicating the presence of noisy files further down the ranking lists. Table \ref{tab:fbr_all} reveals that the performance increases up to a depth of 100 and then starts to decrease, implying that 100 is an optimal reranking depth. We also observe a notable increase in R@100 when the reranking depth is 1,000, suggesting that the CommitReranker could serve as a viable intermediate reranker. However, as noted in Table \ref{tab:query_times}, the CommitReranker is quite inefficient, likely because it considers all commits retrieved by BM25 for a particular file as potential candidates, resulting in a time complexity of $O(\text{depth} \times \text{num\_commits})$ per query.


\paragraph{CodeReranker} demonstrates significantly better performance compared to the CommitReranker, matching the full pipeline's performance at a reranking depth of 100. This suggests that the intermediate CommitReranker may not be as useful. Figure \ref{fig:fbr_non_oracle} also shows a peak at a reranking depth of 100, indicating a bias in the training data creation process. However, the peak is much higher for the CodeReranker compared to the CommitReranker across all metrics, even when trained with less data. This implies that the line-diff-matching data curation approach was quite effective and that scoring with \texttt{maxp} across all passages aligns with intuitive understanding. As the reranking depth increases, the performance of the CodeReranker plateaus. Table \ref{tab:query_times} reveals that the CodeReranker is also quite slow, with a time complexity of $O(\text{rerank\_depth} \times \text{num\_tokens\_per\_file})$ per query, as it needs to process all tokens in each file up to the specified reranking depth.

\paragraph{Peaks at Depth=100}
Figure \ref{fig:fbr_non_oracle} reveals that both the CommitReranker and CodeReranker exhibit increasing performance metrics up to a depth of 100, followed by a drop in metrics. This behavior could be attributed to the training data curation process (Sections \ref{sec:bertrr} and \ref{sec:coderr}), where we consider at most 10 positive and 10 negative examples. Finding these examples might require exploring around 100 documents in the retrieved BM25 results, potentially biasing the models to focus on the top 100 documents and neglect those further down the list. Consequently, the models may not be well-equipped to handle documents (both code patches and commit messages) beyond a depth of 100. To address this issue, we suggest including random easy negatives from lower ranks instead of solely relying on hard negatives ordered by BM25 retrieval.
\paragraph{Full Pipeline Performance}
The full pipeline performs well but does not significantly outperform the CodeReranker at a depth of 100, despite taking 7 times longer to execute. This suggests that a considerable amount of computation may be unnecessary, and the pipeline is examining documents (commits and code patches) that do not substantially contribute to the final results. Developing better techniques to quickly filter out irrelevant documents could be a promising direction for future research.
\paragraph{Overall Non-Oracle Results}
Despite the challenges, our approach achieves a surprisingly good MRR of 0.43 and P@1 of 0.3 across a diverse set of queries implying that for 30\% of the test queries, the first-ranked file is relevant, and on average, one out of every three top-ranked files is relevant. R@10 is also half meaning 1 to 2 out of the three files are easily retrieved. These results demonstrate that leveraging commit messages and cross-training rerankers on the commit history of other repositories is indeed a fruitful approach. However, the remaining files remain a harder problem being much further down the rankings (100-500). 

\paragraph{Oracle Results}
Figure \ref{fig:fbr_oracle} highlights substantial improvements from the pre to post-reranking metrics for both the CommitReranker and CodeReranker, particularly for the CodeReranker at a depth of 100. These results validate the effectiveness of our approach and suggest that if we can sufficiently improve Recall@100 in the current setting, we can achieve near-perfect MRR and P@1. 

\paragraph{CommitReranker Underperformance}
We strongly suspect that the heuristic used for calculating the label for a given training query and retrieved commit message, as described in Section \ref{sec:bertrr}, is the primary reason for the CommitReranker's underperformance. Surprisingly, training on the same data used for the CodeReranker (diff-line-matching) does not yield better performance. This may imply that commit messages are still too noisy to extract more semantic relevance signals, and focusing on source code could be the most promising way forward.

\section{Conclusion}

In this work, we introduced a multi-stage reranking system for repository-level code search that leverages the rich context available in commit histories to identify relevant files for a given query. By combining traditional IR techniques like BM25 with neural reranking using CodeBERT, our system learns to prioritize files based on the semantic similarity of commit messages and the relevance of code changes. Extensive experiments on a diverse set of real-world repositories demonstrate the effectiveness of our approach, with significant improvements over the BM25 baseline.

Key findings include the importance of cross-repository training for generalizability, the superior performance of the CodeReranker compared to the CommitReranker, and that leveraging commit histories is a viable approach. Despite challenges in computational efficiency and ranking consistency, our approach shows promise in enhancing the contextual understanding of large language models for complex coding tasks. 

Future research directions include scaling up the system with more diverse training data, exploring dense embeddings for code file retrieval, better labelling heuristics for CommitReranker and multi-relevance training strategies for improving CodeReranker. Developing better intermediate rerankers and increasing R@100 is the most pertinent problem, closely followed by improving computational efficiency (perhaps quickly filtering out irrelevant documents). Finally, just narrowing data to Github issues which are much less noisy than commits can also be a viable option. 

By providing a concise set of highly relevant files as enriched context, our system has the potential to significantly improve the quality and accuracy of code generation and bug fixing by large language models, ultimately streamlining software development and maintenance processes.

\clearpage
\bibliography{acl2023}
\bibliographystyle{acl_natbib}

\appendix
\section{Appendix}
\label{sec:appendix}
\subsection{Explicit definition of evaluation metrics}
Given a test query $q_t$ for some test commit $c_t$ and a list of actual modified files $F_t$, our system's retrieved list of files $F_r$, and a relevance list $R$ where $R[i]=1$ if file $F_r[i]$ is in $F_t$.

\begin{itemize}
    \item $P@k = \frac{1}{k}\sum_{i=1}^{k}R[i]$
    \item $MRR = 1/(j+1)$ where $j$ is the index of first relevant file in $F_p$ (if none, then 0)
    \item $Recall@K = \frac{1}{|F_a|}\sum_{i=1}^{k}R[i]$
    \item Average Precision $(AP) = \frac{\sum_{j \in \mathcal{X}}P@j}{\sum_{i=1}^{k}R[i]}$, $\mathcal{X}$ is list of indices $i$ where $R[i] = 1$ and macro-averaged across $|Q|$ test queries, $MAP = \frac{AP}{|Q|}$.
\end{itemize}

\subsection{Query Modification Prompt}
\label{app:gpt4_prompt}
The following prompt was provided to GPT4 to modify commit messages into short queries: 


{\small\ttfamily
\vspace{5pt}
You are a professional software developer who is given a very specific task. You will be given commit messages solving one or more problems from any big open-source Github repository, and your task is to identify those core problem(s) that this particular commit is trying to solve. With that information, you need to write a short description of problem itself in the style of a Github issue which would have existed *before* this change was committed. In essence you are trying to reconstruct what potential bugs led to certain commits happening.  Do not mention any information about the solution in the commit (as that would be cheating). Just what the potential issue or problem could have been that led to this commit being needed.

Do not mention the names or contact information of any people involved in the commits (like authors or reviewers). Do not start responding with any description or titles, just start the issue. Limit your response to at most 2 lines. Just think of yourself as a developer who encountered a bug in a hurry and has to write 2 lines which captures as much information about the problem as possible. And remember do NOT leak any details of the solution in the issue message.

For example, given the commit message below:
'[Fizz] Fix for failing id overwrites for postpone (\#27684)
When we postpone during a render we inject a new segment synchronously which we postpone. That gets assigned an ID so we can refer to it immediately in the postponed state.

When we do that, the parent segment may complete later even though it's also synchronous. If that ends up not having any content in it, it'll inline into the child and that will override the child's segment id which is not correct since it was already assigned one.

To fix this, we simply opt-out of the optimization in that case which is unfortunate because we'll generate many more unnecessary empty segments. So we should come up with a new strategy for segment id assignment but this fixes the bug.

Co-authored-by: Josh Story <story@hey.com>'

I want you to respond similar to: 'Synchronous render with postponed segments results in incorrect segment ID overrides, causing empty segments to be generated unnecessarily.'

\vspace{5pt}
}

The modified queries for all repositories are available at \url{https://github.com/Siddharth-Gandhi/ds/tree/main/gold}. In total, it costs \$120 in OpenAI API credits to modify around 6000 queries ((500 train queries + 100 test queries) * 10 repositories). Table \ref{tab:add_transformed_query_examples} contains additional examples of transformed queries for one sample commit from various repositories.

\begin{table*}[h]
\centering
\begin{tabular}{@{}p{0.45\textwidth}p{0.45\textwidth}@{}}

\hline

\textbf{\small\ttfamily Original Commit Message} & \textbf{\small\ttfamily GPT-4 Transformed Short Query} 
\\ \hline

\tiny\ttfamily
[facebook\_react] Extract queueing logic into shared functions (\#22452) As a follow up to \#22445, this extracts the queueing logic that is shared between dispatchSetState and dispatchReducerAction into separate functions. It likely doesn't save any bytes since these will get inlined, anyway, but it does make the flow a bit easier to follow. &
\tiny\ttfamily
There is repeated queueing logic in 'dispatchSetState' and 'dispatchReducerAction', making the code difficult to follow.

\\
\tiny\ttfamily
[angular\_angular] fix(ivy): classes should not mess up matching for bound dir attributes (\#30002) Previously, we had a bug where directive matching could fail if the directive attribute was bound and followed a certain number of classes. This is because in the matching logic, we were treating classes like normal attributes. We should instead be skipping classes in the attribute matching logic. Otherwise classes will match for directives with attribute selectors, and as we are iterating through them in twos (when they are stored as name-only, not in name-value pairs), it may throw off directive matching for any bound attributes that come after. This commit changes the directive matching logic to skip classes altogether. PR Close \#30002 &
\tiny\ttfamily
Directive matching fails when the directive attribute is bound and follows a certain number of classes, as classes are mistakenly treated as normal attributes in the matching logic.

\\
\tiny\ttfamily
[apache\_kafka] KAFKA-13396: Allow create topic without partition/replicaFactor (\#11429) {[}KIP-464{]}(https://cwiki.apache.org/confluence/display/KAFKA/ KIP-464\%3A+Defaults+for+AdminClient\%23createTopic) (PR: https://github.com/apache/kafka/pull/6728) made it possible to create topics without passing partition count and/or replica factor when using the admin client. We incorrectly disallowed this via https://github.com/apache/kafka/pull/10457 while trying to ensure validation was consistent between ZK and the admin client (in this case the inconsistency was intentional). Fix this regression and add tests for the command lines in quick start (i.e. create topic and describe topic) to make sure it won't be broken in the future. Reviewers: Lee Dongjin \textless{}dongjin@apache.org\textgreater{}, Ismael Juma \textless{}ismael@juma.me.uk\textgreater{} & 
\tiny\ttfamily
Inconsistency between validation in ZK and the admin client after applying KIP-464, resulting in inability to create topics without passing partition count and/or replica factor.

\\
\tiny\ttfamily
[apache\_spark] [SPARK-29612][SQL] ALTER TABLE (RECOVER PARTITIONS) should look up catalog/table like v2 commands \#\#\# What changes were proposed in this pull request? Add AlterTableRecoverPartitionsStatement and make ALTER TABLE ... RECOVER PARTITIONS go through the same catalog/table resolution framework of v2 commands. \#\#\# Why are the changes needed? It's important to make all the commands have the same table resolution behavior, to avoid confusing end-users. e.g. ``` USE my\_catalog DESC t // success and describe the table t from my\_catalog ALTER TABLE t RECOVER PARTITIONS // report table not found as there is no table t in the session catalog ``` \#\#\# Does this PR introduce any user-facing change? Yes. When running ALTER TABLE ... RECOVER PARTITIONS Spark fails the command if the current catalog is set to a v2 catalog, or the table name specified a v2 catalog. \#\#\# How was this patch tested? Unit tests. Closes \#26269 from huaxingao/spark-29612. Authored-by: Huaxin Gao <huaxing@us.ibm.com> Signed-off-by: Wenchen Fan <wenchen@databricks.com> & 
\tiny\ttfamily
The current implementation of the "ALTER TABLE ... RECOVER PARTITIONS" command in Spark leads to confusion for users when used with catalog/table resolution. When performed under a catalog (for instance, 'my\_catalog'), the table resolution behaviour is inconsistent. Successful DESCRIBE commands for a table in that catalog are followed by failure in the ALTER TABLE command because it does not find the table in the session catalog. Consequently, there is a need for a fix to ensure uniform table resolution behaviour in all commands, avoiding confusion for end users.

\\
\tiny\ttfamily
[django\_django] Fixed \#26940 -- Removed makemessages from no\_settings\_commands whitelist As makemessages uses several settings for proper run (FILE\_CHARSET, LOCALE\_PATHS, MEDIA\_ROOT, and STATIC\_ROOT), we should require settings configuration for this command. &
\tiny\ttfamily
'makemessages' command is running without required settings configuration (FILE\_CHARSET, LOCALE\_PATHS, MEDIA\_ROOT, STATIC\_ROOT), causing incorrect execution results.

\\
\tiny\ttfamily
[julialang\_julia] Fix type of allocated array when broadcasting type unstable function (\#37028) We need to call similar on the `Broadcasted` object rather than on dest array. Otherwise the `BroadcastStyle` isn't taken into account when allocating new array due to function returning elements of different types. &
\tiny\ttfamily
Broadcasted object does not take into account the BroadcastStyle when allocating new arrays, resulting in type mismatches returned by functions.

\\
\tiny\ttfamily
[redis\_redis] Fix active expire division by zero. Likely fix \#6723. This is what happens AFAIK: we enter the main loop where we expire stuff until a given percentage of keys is still found to be logically expired. There are however other potential exit conditions. However the "sampled" variable is not always incremented inside the loop, because we may found no valid slot as we scan the hash table, but just NULLs ad dict entries. So when the do/while loop condition is triggered at the end, we do (expired*100/sampled), dividing by zero if we sampled 0 keys. &
\tiny\ttfamily
Main loop for key expiry can potentially lead to division by zero error when no valid slots are found during a hash table scan, resulting in no keys being sampled.
\\ \hline
\end{tabular}
\caption{Additional examples of transformed queries for one sample commit from various repositories.}
\label{tab:add_transformed_query_examples}
\end{table*}

\begin{table*}
\centering
\ttfamily
\tiny
\subsection*{\ttfamily{Repository: apache\_kafka}}
\begin{tabular}{|l|r|r|r|r|r|r|r|r|r|r|r|r|r|}
\hline
\textbf{Type} & \textbf{Depth} & \textbf{MAP} & \textbf{MRR} & \textbf{P@1} & \textbf{P@5} & \textbf{P@10} & \textbf{P@20} & \textbf{R@1} & \textbf{R@10} & \textbf{R@20} & \textbf{R@100} & \textbf{R@500} & \textbf{R@1000} \\
\hline
\textbf{BM25} & \textbf{} & 0.142 & 0.273 & 0.180 & 0.102 & 0.087 & 0.066 & 0.052 & 0.206 & 0.314 & 0.590 & 0.814 & 0.889 \\
\cline{1-14}
\multirow[t]{4}{*}{\textbf{CommitReranker}} & \textbf{100} & 0.226 & 0.368 & 0.230 & 0.170 & 0.131 & 0.093 & 0.074 & 0.365 & 0.460 & 0.590 & 0.814 & 0.889 \\
\textbf{} & \textbf{250} & 0.212 & 0.326 & 0.200 & 0.146 & 0.117 & 0.092 & 0.066 & 0.305 & 0.457 & 0.638 & 0.814 & 0.889 \\
\textbf{} & \textbf{500} & 0.198 & 0.295 & 0.170 & 0.130 & 0.105 & 0.084 & 0.062 & 0.291 & 0.410 & 0.664 & 0.814 & 0.889 \\
\textbf{} & \textbf{1000} & 0.180 & 0.268 & 0.160 & 0.112 & 0.090 & 0.074 & 0.060 & 0.263 & 0.367 & 0.647 & 0.824 & 0.889 \\
\cline{1-14}
\multirow[t]{4}{*}{\textbf{CodeReranker}} & \textbf{100} & 0.272 & 0.453 & 0.340 & 0.186 & 0.135 & 0.090 & 0.117 & 0.371 & 0.450 & 0.590 & 0.814 & 0.889 \\
\textbf{} & \textbf{250} & 0.247 & 0.452 & 0.340 & 0.184 & 0.140 & 0.097 & 0.092 & 0.368 & 0.459 & 0.672 & 0.814 & 0.889 \\
\textbf{} & \textbf{500} & 0.235 & 0.422 & 0.310 & 0.162 & 0.135 & 0.096 & 0.096 & 0.347 & 0.453 & 0.714 & 0.814 & 0.889 \\
\textbf{} & \textbf{1000} & 0.230 & 0.415 & 0.320 & 0.156 & 0.117 & 0.086 & 0.104 & 0.312 & 0.421 & 0.683 & 0.839 & 0.889 \\
\cline{1-14}
\textbf{Full Pipeline} & \textbf{} & 0.277 & 0.509 & 0.400 & 0.218 & 0.152 & 0.102 & 0.108 & 0.380 & 0.470 & 0.647 & 0.824 & 0.889 \\
\cline{1-14}
\hline
\end{tabular}

\subsection*{\ttfamily{Repository: julia\_julia}}
\begin{tabular}{|l|r|r|r|r|r|r|r|r|r|r|r|r|r|}
\hline
\textbf{Type} & \textbf{Depth} & \textbf{MAP} & \textbf{MRR} & \textbf{P@1} & \textbf{P@5} & \textbf{P@10} & \textbf{P@20} & \textbf{R@1} & \textbf{R@10} & \textbf{R@20} & \textbf{R@100} & \textbf{R@500} & \textbf{R@1000} \\
\hline
\textbf{BM25} & \textbf{} & 0.171 & 0.234 & 0.160 & 0.074 & 0.058 & 0.036 & 0.083 & 0.277 & 0.332 & 0.718 & 0.978 & 0.995 \\
\cline{1-14}
\multirow[t]{4}{*}{\textbf{CommitReranker}} & \textbf{100} & 0.292 & 0.377 & 0.250 & 0.142 & 0.098 & 0.060 & 0.137 & 0.481 & 0.577 & 0.718 & 0.978 & 0.995 \\
\textbf{} & \textbf{250} & 0.269 & 0.355 & 0.230 & 0.132 & 0.087 & 0.058 & 0.124 & 0.453 & 0.577 & 0.817 & 0.978 & 0.995 \\
\textbf{} & \textbf{500} & 0.251 & 0.335 & 0.210 & 0.118 & 0.083 & 0.057 & 0.109 & 0.433 & 0.558 & 0.819 & 0.978 & 0.995 \\
\textbf{} & \textbf{1000} & 0.248 & 0.332 & 0.210 & 0.116 & 0.082 & 0.056 & 0.109 & 0.432 & 0.555 & 0.816 & 0.984 & 0.995 \\
\cline{1-14}
\multirow[t]{4}{*}{\textbf{CodeReranker}} & \textbf{100} & 0.276 & 0.363 & 0.240 & 0.132 & 0.089 & 0.057 & 0.139 & 0.418 & 0.551 & 0.718 & 0.978 & 0.995 \\
\textbf{} & \textbf{250} & 0.238 & 0.319 & 0.210 & 0.112 & 0.085 & 0.056 & 0.123 & 0.412 & 0.523 & 0.812 & 0.978 & 0.995 \\
\textbf{} & \textbf{500} & 0.224 & 0.306 & 0.210 & 0.100 & 0.070 & 0.052 & 0.125 & 0.334 & 0.489 & 0.801 & 0.978 & 0.995 \\
\textbf{} & \textbf{1000} & 0.209 & 0.283 & 0.180 & 0.096 & 0.064 & 0.050 & 0.115 & 0.313 & 0.469 & 0.735 & 0.960 & 0.995 \\
\cline{1-14}
\textbf{Full Pipeline} & \textbf{} & 0.283 & 0.362 & 0.230 & 0.138 & 0.099 & 0.064 & 0.132 & 0.489 & 0.610 & 0.816 & 0.984 & 0.995 \\
\cline{1-14}
\hline
\end{tabular}

\subsection*{\ttfamily{Repository: django\_django}}
\begin{tabular}{|l|r|r|r|r|r|r|r|r|r|r|r|r|r|}
\hline
\textbf{Type} & \textbf{Depth} & \textbf{MAP} & \textbf{MRR} & \textbf{P@1} & \textbf{P@5} & \textbf{P@10} & \textbf{P@20} & \textbf{R@1} & \textbf{R@10} & \textbf{R@20} & \textbf{R@100} & \textbf{R@500} & \textbf{R@1000} \\
\hline
\textbf{BM25} & \textbf{} & 0.232 & 0.278 & 0.210 & 0.104 & 0.064 & 0.038 & 0.113 & 0.328 & 0.388 & 0.616 & 0.840 & 0.950 \\
\cline{1-14}
\multirow[t]{4}{*}{\textbf{CommitReranker}} & \textbf{100} & 0.285 & 0.374 & 0.290 & 0.118 & 0.083 & 0.049 & 0.158 & 0.438 & 0.513 & 0.616 & 0.840 & 0.950 \\
\textbf{} & \textbf{250} & 0.272 & 0.343 & 0.260 & 0.120 & 0.073 & 0.050 & 0.143 & 0.376 & 0.528 & 0.763 & 0.840 & 0.950 \\
\textbf{} & \textbf{500} & 0.282 & 0.361 & 0.270 & 0.126 & 0.079 & 0.050 & 0.148 & 0.402 & 0.515 & 0.747 & 0.840 & 0.950 \\
\textbf{} & \textbf{1000} & 0.267 & 0.332 & 0.250 & 0.114 & 0.072 & 0.048 & 0.133 & 0.371 & 0.498 & 0.777 & 0.938 & 0.950 \\
\cline{1-14}
\multirow[t]{4}{*}{\textbf{CodeReranker}} & \textbf{100} & 0.336 & 0.439 & 0.310 & 0.162 & 0.100 & 0.056 & 0.178 & 0.522 & 0.579 & 0.616 & 0.840 & 0.950 \\
\textbf{} & \textbf{250} & 0.336 & 0.432 & 0.290 & 0.152 & 0.102 & 0.062 & 0.169 & 0.546 & 0.649 & 0.762 & 0.840 & 0.950 \\
\textbf{} & \textbf{500} & 0.313 & 0.400 & 0.250 & 0.142 & 0.090 & 0.058 & 0.149 & 0.491 & 0.628 & 0.803 & 0.840 & 0.950 \\
\textbf{} & \textbf{1000} & 0.307 & 0.397 & 0.240 & 0.146 & 0.087 & 0.058 & 0.128 & 0.477 & 0.633 & 0.852 & 0.950 & 0.950 \\
\cline{1-14}
\textbf{Full Pipeline} & \textbf{} & 0.387 & 0.482 & 0.330 & 0.180 & 0.116 & 0.068 & 0.189 & 0.624 & 0.725 & 0.777 & 0.938 & 0.950 \\
\cline{1-14}
\hline
\end{tabular}

\subsection*{\ttfamily{Repository: redis\_redis}}
\begin{tabular}{|l|r|r|r|r|r|r|r|r|r|r|r|r|r|}
\hline
\textbf{Type} & \textbf{Depth} & \textbf{MAP} & \textbf{MRR} & \textbf{P@1} & \textbf{P@5} & \textbf{P@10} & \textbf{P@20} & \textbf{R@1} & \textbf{R@10} & \textbf{R@20} & \textbf{R@100} & \textbf{R@500} & \textbf{R@1000} \\
\hline
\textbf{BM25} & \textbf{} & 0.272 & 0.343 & 0.190 & 0.132 & 0.115 & 0.074 & 0.106 & 0.540 & 0.658 & 0.904 & 0.996 & 0.998 \\
\cline{1-14}
\multirow[t]{4}{*}{\textbf{CommitReranker}} & \textbf{100} & 0.409 & 0.533 & 0.400 & 0.190 & 0.137 & 0.086 & 0.219 & 0.635 & 0.781 & 0.904 & 0.996 & 0.998 \\
\textbf{} & \textbf{250} & 0.397 & 0.514 & 0.380 & 0.190 & 0.128 & 0.082 & 0.211 & 0.602 & 0.749 & 0.935 & 0.996 & 0.998 \\
\textbf{} & \textbf{500} & 0.392 & 0.509 & 0.370 & 0.188 & 0.128 & 0.082 & 0.201 & 0.601 & 0.742 & 0.948 & 0.996 & 0.998 \\
\textbf{} & \textbf{1000} & 0.391 & 0.509 & 0.370 & 0.188 & 0.127 & 0.081 & 0.201 & 0.596 & 0.732 & 0.948 & 0.992 & 0.998 \\
\cline{1-14}
\multirow[t]{4}{*}{\textbf{CodeReranker}} & \textbf{100} & 0.467 & 0.590 & 0.450 & 0.218 & 0.145 & 0.091 & 0.244 & 0.664 & 0.819 & 0.904 & 0.996 & 0.998 \\
\textbf{} & \textbf{250} & 0.435 & 0.540 & 0.400 & 0.200 & 0.138 & 0.082 & 0.230 & 0.632 & 0.751 & 0.930 & 0.996 & 0.998 \\
\textbf{} & \textbf{500} & 0.411 & 0.513 & 0.370 & 0.196 & 0.129 & 0.078 & 0.218 & 0.599 & 0.694 & 0.953 & 0.996 & 0.998 \\
\textbf{} & \textbf{1000} & 0.392 & 0.490 & 0.340 & 0.188 & 0.123 & 0.077 & 0.201 & 0.577 & 0.681 & 0.931 & 0.992 & 0.998 \\
\cline{1-14}
\textbf{Full Pipeline} & \textbf{} & 0.474 & 0.584 & 0.450 & 0.214 & 0.145 & 0.089 & 0.259 & 0.673 & 0.808 & 0.948 & 0.992 & 0.998 \\
\cline{1-14}
\hline
\end{tabular}

\subsection*{\ttfamily{Repository: apache\_spark}}
\begin{tabular}{|l|r|r|r|r|r|r|r|r|r|r|r|r|r|}
\hline
\textbf{Type} & \textbf{Depth} & \textbf{MAP} & \textbf{MRR} & \textbf{P@1} & \textbf{P@5} & \textbf{P@10} & \textbf{P@20} & \textbf{R@1} & \textbf{R@10} & \textbf{R@20} & \textbf{R@100} & \textbf{R@500} & \textbf{R@1000} \\
\hline
\textbf{BM25} & \textbf{} & 0.088 & 0.116 & 0.070 & 0.054 & 0.038 & 0.025 & 0.023 & 0.116 & 0.135 & 0.415 & 0.790 & 0.881 \\
\cline{1-14}
\multirow[t]{4}{*}{\textbf{CommitReranker}} & \textbf{100} & 0.158 & 0.253 & 0.170 & 0.104 & 0.077 & 0.051 & 0.054 & 0.232 & 0.280 & 0.415 & 0.790 & 0.881 \\
\textbf{} & \textbf{250} & 0.200 & 0.284 & 0.190 & 0.104 & 0.079 & 0.057 & 0.093 & 0.268 & 0.374 & 0.610 & 0.790 & 0.881 \\
\textbf{} & \textbf{500} & 0.156 & 0.226 & 0.130 & 0.072 & 0.067 & 0.051 & 0.060 & 0.238 & 0.336 & 0.574 & 0.790 & 0.881 \\
\textbf{} & \textbf{1000} & 0.144 & 0.208 & 0.120 & 0.068 & 0.061 & 0.040 & 0.058 & 0.218 & 0.274 & 0.559 & 0.812 & 0.881 \\
\cline{1-14}
\multirow[t]{4}{*}{\textbf{CodeReranker}} & \textbf{100} & 0.218 & 0.332 & 0.230 & 0.130 & 0.090 & 0.064 & 0.096 & 0.283 & 0.356 & 0.415 & 0.790 & 0.881 \\
\textbf{} & \textbf{250} & 0.254 & 0.379 & 0.250 & 0.158 & 0.112 & 0.074 & 0.102 & 0.406 & 0.481 & 0.662 & 0.790 & 0.881 \\
\textbf{} & \textbf{500} & 0.216 & 0.322 & 0.180 & 0.132 & 0.097 & 0.068 & 0.071 & 0.361 & 0.461 & 0.687 & 0.790 & 0.881 \\
\textbf{} & \textbf{1000} & 0.207 & 0.317 & 0.190 & 0.118 & 0.090 & 0.066 & 0.078 & 0.326 & 0.463 & 0.682 & 0.849 & 0.881 \\
\cline{1-14}
\textbf{Full Pipeline} & \textbf{} & 0.256 & 0.394 & 0.270 & 0.166 & 0.107 & 0.066 & 0.106 & 0.376 & 0.440 & 0.559 & 0.812 & 0.881 \\
\cline{1-14}
\hline
\end{tabular}

\subsection*{\ttfamily{Repository: angular\_angular}}
\begin{tabular}{|l|r|r|r|r|r|r|r|r|r|r|r|r|r|}
\hline
\textbf{Type} & \textbf{Depth} & \textbf{MAP} & \textbf{MRR} & \textbf{P@1} & \textbf{P@5} & \textbf{P@10} & \textbf{P@20} & \textbf{R@1} & \textbf{R@10} & \textbf{R@20} & \textbf{R@100} & \textbf{R@500} & \textbf{R@1000} \\
\hline
\textbf{BM25} & \textbf{} & 0.116 & 0.165 & 0.080 & 0.068 & 0.059 & 0.042 & 0.027 & 0.197 & 0.266 & 0.452 & 0.708 & 0.803 \\
\cline{1-14}
\multirow[t]{4}{*}{\textbf{CommitReranker}} & \textbf{100} & 0.206 & 0.294 & 0.210 & 0.106 & 0.078 & 0.053 & 0.104 & 0.297 & 0.346 & 0.452 & 0.708 & 0.803 \\
\textbf{} & \textbf{250} & 0.206 & 0.286 & 0.200 & 0.108 & 0.080 & 0.060 & 0.082 & 0.297 & 0.399 & 0.596 & 0.708 & 0.803 \\
\textbf{} & \textbf{500} & 0.193 & 0.274 & 0.190 & 0.098 & 0.072 & 0.060 & 0.078 & 0.270 & 0.394 & 0.606 & 0.708 & 0.803 \\
\textbf{} & \textbf{1000} & 0.187 & 0.263 & 0.180 & 0.100 & 0.074 & 0.057 & 0.068 & 0.289 & 0.379 & 0.625 & 0.776 & 0.803 \\
\cline{1-14}
\multirow[t]{4}{*}{\textbf{CodeReranker}} & \textbf{100} & 0.181 & 0.262 & 0.150 & 0.112 & 0.084 & 0.054 & 0.081 & 0.304 & 0.373 & 0.452 & 0.708 & 0.803 \\
\textbf{} & \textbf{250} & 0.174 & 0.250 & 0.120 & 0.104 & 0.089 & 0.063 & 0.051 & 0.330 & 0.430 & 0.590 & 0.708 & 0.803 \\
\textbf{} & \textbf{500} & 0.167 & 0.240 & 0.100 & 0.108 & 0.085 & 0.063 & 0.035 & 0.311 & 0.427 & 0.622 & 0.708 & 0.803 \\
\textbf{} & \textbf{1000} & 0.154 & 0.228 & 0.100 & 0.100 & 0.073 & 0.056 & 0.034 & 0.276 & 0.386 & 0.647 & 0.774 & 0.803 \\
\cline{1-14}
\textbf{Full Pipeline} & \textbf{} & 0.202 & 0.271 & 0.130 & 0.110 & 0.096 & 0.067 & 0.071 & 0.352 & 0.471 & 0.625 & 0.776 & 0.803 \\
\cline{1-14}
\hline
\end{tabular}

\subsection*{\ttfamily{Repository: facebook\_react}}
\begin{tabular}{|l|r|r|r|r|r|r|r|r|r|r|r|r|r|}
\hline
\textbf{Type} & \textbf{Depth} & \textbf{MAP} & \textbf{MRR} & \textbf{P@1} & \textbf{P@5} & \textbf{P@10} & \textbf{P@20} & \textbf{R@1} & \textbf{R@10} & \textbf{R@20} & \textbf{R@100} & \textbf{R@500} & \textbf{R@1000} \\
\hline
\textbf{BM25} & \textbf{} & 0.119 & 0.180 & 0.080 & 0.094 & 0.077 & 0.056 & 0.022 & 0.189 & 0.260 & 0.557 & 0.797 & 0.951 \\
\cline{1-14}
\multirow[t]{4}{*}{\textbf{CommitReranker}} & \textbf{100} & 0.187 & 0.330 & 0.200 & 0.168 & 0.114 & 0.076 & 0.053 & 0.280 & 0.373 & 0.557 & 0.797 & 0.951 \\
\textbf{} & \textbf{250} & 0.186 & 0.328 & 0.220 & 0.140 & 0.105 & 0.070 & 0.069 & 0.276 & 0.368 & 0.565 & 0.797 & 0.951 \\
\textbf{} & \textbf{500} & 0.184 & 0.315 & 0.200 & 0.126 & 0.101 & 0.068 & 0.066 & 0.279 & 0.356 & 0.559 & 0.797 & 0.951 \\
\textbf{} & \textbf{1000} & 0.185 & 0.306 & 0.180 & 0.130 & 0.100 & 0.068 & 0.061 & 0.293 & 0.369 & 0.590 & 0.874 & 0.951 \\
\cline{1-14}
\multirow[t]{4}{*}{\textbf{CodeReranker}} & \textbf{100} & 0.267 & 0.420 & 0.300 & 0.202 & 0.144 & 0.094 & 0.097 & 0.382 & 0.465 & 0.557 & 0.797 & 0.951 \\
\textbf{} & \textbf{250} & 0.240 & 0.355 & 0.200 & 0.180 & 0.141 & 0.096 & 0.076 & 0.374 & 0.477 & 0.630 & 0.797 & 0.951 \\
\textbf{} & \textbf{500} & 0.236 & 0.351 & 0.210 & 0.156 & 0.127 & 0.098 & 0.081 & 0.352 & 0.514 & 0.683 & 0.797 & 0.951 \\
\textbf{} & \textbf{1000} & 0.225 & 0.334 & 0.200 & 0.146 & 0.117 & 0.089 & 0.076 & 0.332 & 0.470 & 0.740 & 0.901 & 0.951 \\
\cline{1-14}
\textbf{Full Pipeline} & \textbf{} & 0.270 & 0.434 & 0.280 & 0.202 & 0.147 & 0.094 & 0.109 & 0.393 & 0.480 & 0.590 & 0.874 & 0.951 \\
\cline{1-14}
\hline
\end{tabular}
\caption{Results for various configuartions per repository}
\label{tab:per_repo_results}
\end{table*}

\subsection{Algorithm for getting FIDs}
\label{app:fid}
The algorithm is described at this link \url{https://github.com/Siddharth-Gandhi/ds/blob/boston/notebooks/git_map.ipynb}. The FID mappings for the various repositories tested are available at \url{https://github.com/Siddharth-Gandhi/ds/tree/boston/fids/v4}.


\subsection{Results per Repository}
\label{app:per_repo_results}
Table \ref{tab:per_repo_results} details all of the results for BM25, various rerank depths of CommitReranker and CodeReranker and Full Pipeline (BM25 $\rightarrow$ CommitReranker @ 1000 $\rightarrow$ CodeReranker @ 100) configurations.
\end{document}